# Chaotic mode-competition dynamics in a multimode semiconductor laser with optical feedback and injection

Ryugo Iwami,[1,*] Kazutaka Kanno,[1] and Atsushi Uchida[1,**]

[1]*Department of Information and Computer Sciences, Saitama University,*
*255 Shimo-okubo, Sakura-ku, Saitama City, Saitama, 338-8570, Japan.*
*\*r.iwami.692@ms.saitama-u.ac.jp*
*\*\*auchida@mail.saitama-u.ac.jp*

**Abstract:** Photonic computing is attracting increasing interest to accelerate information processing in machine learning applications. The mode-competition dynamics of multimode semiconductor lasers is useful for solving the multi-armed bandit problem in reinforcement learning for computing applications. In this study, we numerically evaluate the chaotic mode-competition dynamics in a multimode semiconductor laser with optical feedback and injection. We observe the chaotic mode-competition dynamics among the longitudinal modes and control them by injecting an external optical signal into one of the longitudinal modes. We define the dominant mode as the mode with the maximum intensity; the dominant-mode ratio for the injected mode increases as the optical injection strength increases. We find that the characteristics of the dominant mode ratio in terms of the optical injection strength are different among the modes owing to the different optical feedback phases. We propose a control technique for the characteristics of the dominant mode ratio by precisely tuning the initial optical frequency detuning between the optical injection signal and injected mode. We also evaluate the relationship between the region for the large dominant mode ratio and injection locking range. The region for the large dominant mode ratio does not correspond to the injection-locking range. This discrepancy results from the complex mode-competition dynamics in multimode semiconductor lasers with both optical feedback and injection. This control technique of chaotic mode-competition dynamics in multimode lasers is promising for applications in reinforcement learning and reservoir computing as photonic artificial intelligence.

## 1. Introduction

Photonic accelerators, which accelerate specific information processing using light, have been widely studied as novel computation technologies in the post-Moore era [1]. In photonic accelerators, photonic technologies have been used for artificial intelligence (AI) such as photonic neural networks [2], coherent Ising machines [3], photonic reservoir computing [4–6], and photonic decision-making [7–10]. The use of semiconductor laser dynamics in the development of photonic AI technologies is promising.

Controlling chaos has been applied to many nonlinear dynamical systems, which stabilize chaotic outputs into a steady state or periodic output [11,12]. The concept of controlling chaos, referred to as the OGY method, was first proposed by Ott et al. [11], and multiple studies on controlling chaos have been reported [12,13]. Controlling chaos in lasers has been experimentally achieved [14], and chaotic oscillations in lasers can be stabilized into steady-state or periodic outputs. Techniques for controlling chaos in lasers have been applied for the stabilization of chaos into high periodic oscillations [15], suppression of relative intensity noise [16], and dynamic associative memory [17]. The diversity of controlling chaos could be useful for achieving advanced information processing in AI because its complex behavior and controllability allow spontaneous exploration and exploitation functions in reinforcement learning [18]. Thus, the evaluation of the controlling chaos in lasers is of great significance.

Numerous studies have been conducted on the dynamics of semiconductor lasers in the presence of either optical feedback [19–22] or optical injection [23–25]. Various nonlinear dynamics and bifurcation phenomena have been reported in literature. However, studies on the dynamics of semiconductor lasers in the presence of both optical feedback and injection are limited compared to those on semiconductor lasers with either of them. Particularly, bandwidth enhancement of chaotic oscillations has been studied in semiconductor lasers with optical feedback and injection [26]. Additionally, semiconductor lasers with optical feedback and injection have been used for photonic reservoir computing [27–29]. The performance of photonic reservoir computing in these systems is sensitive to the phase of the feedback and injection light because a change in the optical phase results in different dynamics [27,28]. The optical feedback phase strongly affects the dynamics of a semiconductor laser in the short external cavity regime [30].

A variety of studies on nonlinear dynamics have been reported for multimode (Fabry–Perot) semiconductor lasers with multiple longitudinal modes. Spontaneous mode hopping is induced by optical injection in a multimode semiconductor laser [31,32]. Modal and total intensity dynamics in the low-frequency fluctuation (LFF) regime have been studied in a multimode semiconductor laser with optical feedback [33–38]. The interaction of the longitudinal modes plays an important role in the LFF of multimode semiconductor lasers, which indicates the existence of an anti-correlated interaction among the modal intensities (called anti-phase dynamics [12,39]). Chaotic antiphase dynamics have been observed experimentally in a multimode semiconductor laser with optical feedback [39]. Moreover, the mode with the maximum intensity (i.e., the dominant mode) competes chaotically [40], and it can be adaptively selected from the chaotic mode-competition dynamics by changing the optical feedback strength or injection current [40].

However, the control of chaotic mode-competition dynamics via optical injection has not yet been studied. A control technique for chaotic mode-competition dynamics in a multimode semiconductor laser can be applied for photonic information processing to solve the multi-armed bandit problem in reinforcement learning [18]. Additionally, multimode semiconductor lasers have been used for photonic reservoir computing to process multiple tasks in parallel using multimode dynamics [29]. Therefore, they are expected to have high potential as photonic accelerators for computing applications.

In this study, we numerically evaluate the chaotic mode-competition dynamics in a multimode semiconductor laser with optical feedback and injection. We introduce a technique for controlling the dominant mode in a multimode laser by injecting optical signals from stable single-mode semiconductor lasers. We also evaluate the relationship between the parameter regions of the large dominant-mode ratio and the injection-locking range.

## 2. Numerical model

Figure 1 shows the schematic of our numerical model for a multimode semiconductor laser with optical feedback and injection. We consider a single-mode semiconductor laser for optical injection. The mode-competition dynamics in a multimode semiconductor laser with optical feedback are controlled by exciting the longitudinal mode $m$ with modal frequency $v_m$ by optical injection from a single-mode semiconductor laser. The optical frequency of the single-mode semiconductor laser is defined as $f_m$, whose frequency is near $v_m$. The initial optical-frequency detuning between the injection light and injected mode is defined as $\Delta f_m = f_m - v_m$.

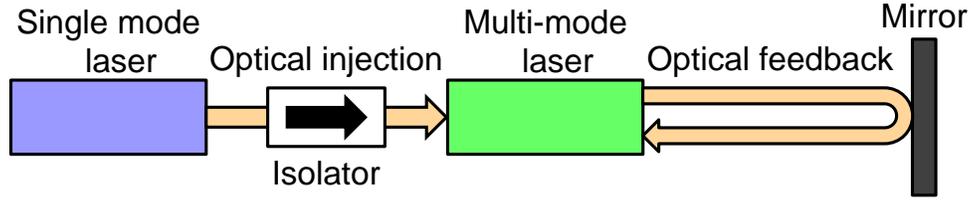

Fig. 1. Schematic of numerical model for multimode semiconductor laser with optical feedback and injection. Light from a single-mode semiconductor laser is injected into a multimode semiconductor laser with optical feedback to control one of the longitudinal modes in the multimode laser.

**Table 1. Parameter values used in numerical simulations.**

| Symbol | Parameter | Value |
| --- | --- | --- |
| $M$ | Number of longitudinal modes | 5 |
| $G_N$ | Gain coefficient of central mode (mode 3) | $8.40 \times 10^{-13}$ m$^3$s$^{-1}$ |
| $N_0$ | Carrier density at transparency | $1.40 \times 10^{24}$ m$^{-3}$ |
| $\alpha$ | Linewidth enhancement factor | 3.0 |
| $\varepsilon$ | Gain saturation coefficient | $2.5 \times 10^{-23}$ |
| $\tau_p$ | Photon lifetime | $1.927 \times 10^{-12}$ s |
| $\tau_s$ | Carrier lifetime | $2.04 \times 10^{-9}$ s |
| $\kappa$ | Optical feedback strength | $4.411 \times 10^{9}$ s$^{-1}$ |
| $\kappa_{inj,m}$ | Optical injection strength for mode $m$ | Variable ($0.0 \sim 15.0 \times 10^9$ s$^{-1}$) |
| $\tau$ | Round-trip time of light in external cavity | $1.001 \times 10^{-8}$ s |
| $L$ | External cavity length (one way) | 1.5 m |
| $J$ | Injection current | $1.11\, J_{th}$ |
| $\Delta\nu$ | Frequency of longitudinal mode spacing | $3.55 \times 10^{10}$ Hz |
| $\nu_c$ | Frequency of central mode | $1.951 \times 10^{14}$ Hz |
| $\Delta\nu_g$ | Frequency width of gain profile | $1.270 \times 10^{13}$ Hz |
| $\Delta f_m$ | Initial optical frequency detuning for mode $m$ | Variable ($-2.0 \times 10^{10} \sim 2.0 \times 10^{10}$ Hz) |
| $A_s$ | Steady-state solution of electric field amplitude of single-mode semiconductor laser | $1.438 \times 10^{10}$ |
| $J_{th} = N_{th}/\tau_s$ | Injection current at lasing threshold | $9.891 \times 10^{32}$ m$^{-3}$s$^{-1}$ |
| $N_{th} = N_0 + 1/(G_N\tau_p)$ | Carrier density at lasing threshold | $2.018 \times 10^{24}$ m$^{-3}$ |

We use a numerical model described by the Lang–Kobayashi equations [12,41,42], which are well-known rate equations for a single-mode semiconductor laser with optical feedback. The Lang–Kobayashi equations can be extended to multiple longitudinal modes [37,40,42]. The effect of optical injection from a single-mode semiconductor laser can be added [12,26,42]. The numerical model of a multimode semiconductor laser with $M$ longitudinal modes under optical injection is described as follows:

$$\frac{dE_m(t)}{dt} = \frac{1-i\alpha}{2}\left\{\frac{G_m[N(t)-N_0]}{1+\varepsilon\sum_{k=1}^{M}|E_k(t)|^2} - \frac{1}{\tau_p}\right\}E_m(t) \\ + \kappa E_m(t-\tau)\exp(i\omega_m\tau) + \kappa_{inj,m}A_s\exp(-i2\pi\Delta f_m t) \quad (1)$$

$$\frac{dN(t)}{dt} = J - \frac{N(t)}{\tau_s} - \sum_{l=1}^{M} \left\{ \frac{G_l[N(t) - N_0]|E_l(t)|^2}{1 + \varepsilon \sum_{k=1}^{M} |E_k(t)|^2} \right\} \quad (2)$$

$$G_m = G_N \left[ 1 - \frac{(v_m - v_{m_c})^2}{\Delta v_g^2} \right] \quad (3)$$

where $E_m(t)$ represents the complex electric field amplitude of longitudinal mode $m$ and $N(t)$ represents the carrier density; $G_m$ represents the gain coefficient of mode $m$; and, $i$ represents the imaginary unit. In this study, we considered five longitudinal modes: $M = 5$. The frequency of mode $m$ ($v_m$) is defined as the mode spacing $\Delta v$, that is, $v_m = v_c + (m - m_c)\Delta v$, where $v_c$ and $m_c$ are the frequency and mode number of the central mode, respectively. Here, $m_c$ is described as $m_c = (M + 1) / 2$ when $M$ is assumed to be an odd number; $m_c$ is set to 3 in this study. The optical angular frequency of the mode $m$ is defined as $\omega_m = 2\pi v_m$. The modal intensity of the laser output was calculated as $I_m(t) = |E_m(t)|^2$. The total intensity of the multimode laser output is obtained by the incoherent sum of the modal intensities, that is, $I_{total}(t) = \Sigma|E_m(t)|^2$. The parameter values used in the numerical simulations are presented in Table 1.

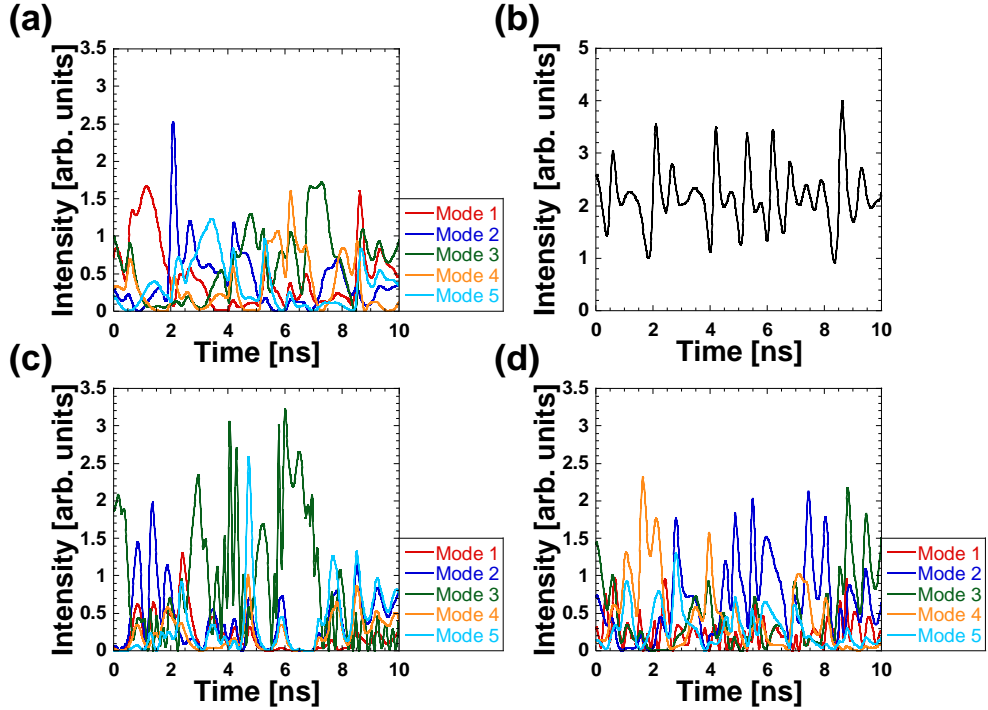

Fig. 2. Temporal waveforms of multimode semiconductor laser with optical feedback and injection. (a) Modal intensities when optical feedback is only applied without optical injection. (b) Total intensity corresponding to (a). (c) Modal intensities when both optical feedback and injection are applied to mode 3 at $\kappa_{inj,3} = 6.0$ ns$^{-1}$. (d) Modal intensities when both optical feedback and injection are applied to mode 1 at $\kappa_{inj,1} = 6.0$ ns$^{-1}$. Initial optical frequency detuning is fixed at $\Delta f_m = -4.0$ GHz in (c) and (d).

## 3. Numerical results

*3.1 Mode-competition dynamics with respect to optical injection*

We evaluated the temporal waveforms with and without optical injection under optical feedback. Figure 2 shows the numerical results for the temporal waveforms of the multimode semiconductor laser. Figure 2(a) shows the five modal intensities when only optical feedback was applied (i.e., no optical injection). The optical feedback strength was set to $\kappa = 4.411$ ns$^{-1}$. In Fig. 2(a), each modal intensity shows chaotic oscillation; however, the oscillation is different for each mode. We define the mode with the maximum intensity as the dominant mode. The dominant mode changes in time and the chaotic mode-competition dynamics occurs. Figure 2(b) shows the total intensity corresponding to Fig. 2(a). The total intensity exhibited chaotic oscillation.

Figure 2(c) shows the modal intensities when the optical injection is applied to mode 3 with $\kappa_{inj,3} = 6.0$ ns$^{-1}$ under optical feedback. Here, the initial optical frequency detuning was fixed at $\Delta f_m = -4.0$ GHz. Mode 3 (green curve) oscillates with a large amplitude owing to the optical injection to mode 3, and the duration of the dominant mode for mode 3 is longer than that for the other modes. In other words, mode 3 was excited by optical injection. Figure 2(d) shows the modal intensities when the optical injection is applied to mode 1 with $\kappa_{inj,1} = 6.0$ ns$^{-1}$ under optical feedback. In Fig. 2(d), the oscillation of mode 1 (red curve) is suppressed compared with that of the other modes, and the duration of the dominant mode for mode 1 is shorter than that of the other modes. From Figs. 2(c) and 2(d), the behaviors of the mode-competition dynamics are different among the injected modes even though the injection strength is set to the same value.

We determined the change in the dominant mode to quantitatively evaluate the mode-competition dynamics when the optical injection strength was changed. The dominant mode ratio is defined as the ratio of the dominant mode of mode $m$ over a long period [40]. The dominant-mode ratio $DMR_m$ for mode $m$ is expressed as follows:

$$DMR_m = \frac{1}{S}\sum_{j=1}^{S} D_m(j) \qquad (4)$$

where $S$ is the total number of sample points corresponding to time length. $D_m(j)$ is 1 if the modal intensity of mode $m$ is the dominant mode (i.e., the maximum intensity among the modes) at the $j$-th sampling point and 0 otherwise.

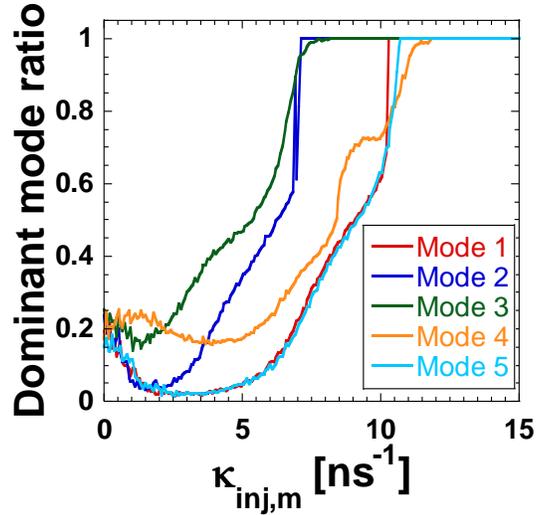

Fig. 3. Comparison of dominant mode ratios of mode $m$ as a function of optical injection strength for mode $m$ when optical injection is applied for only mode $m$ under optical feedback. Initial optical frequency detuning for each mode is $\Delta f_m = -4.0$ GHz.

Figure 3 shows the dominant mode ratio of mode $m$ when an optical injection is applied only to mode $m$ and the optical injection strength for mode $m$ ($\kappa_{inj,m}$) is changed. The initial optical frequency detuning between the injection light and injected mode are set to $\Delta f_m = -4.0$ GHz for all the modes. In Fig. 3, the dominant mode ratio changes differently for each mode; however, the characteristics of the dominant modes for modes 1 and 5 are similar. Particularly, the dominant mode ratio for mode 3 increased as the optical injection strength increased, whereas those of modes 1, 2, and 5 decreased to near zero for a small $\kappa_{inj,m}$. The dominant mode ratio reaches 1 eventually for all modes when the optical injection strength is sufficiently large. A large injection strength results in the growth of the dominant mode; however, the characteristics of the dominant mode ratio in terms of the optical injection strength are different among the modes, as shown in Fig. 3.

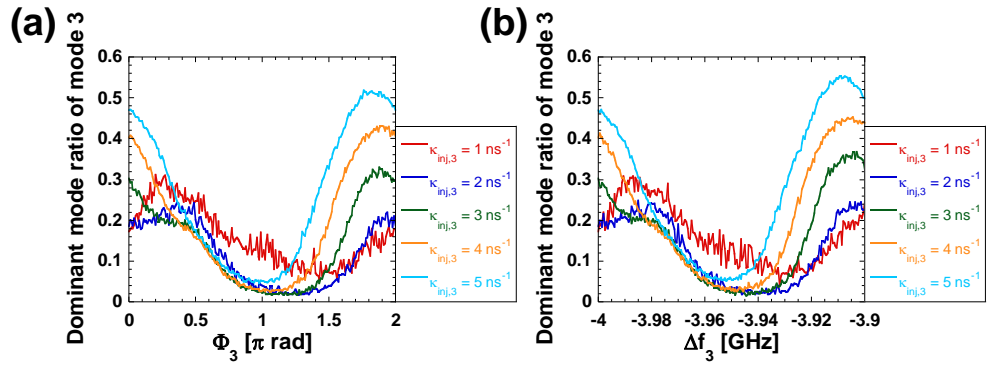

Fig. 4. Comparison of dominant mode ratios of mode 3 when optical injection is applied for mode 3 at different optical injection strengths under optical feedback. Each color corresponds to the optical injection strength to mode 3 ($\kappa_{inj,3}$). Optical injection strength is fixed at $\kappa_{inj,3} = 1.0$ ns$^{-1}$, 2.0 ns$^{-1}$, 3.0 ns$^{-1}$, 4.0 ns$^{-1}$, and 5.0 ns$^{-1}$. (a) Dominant mode ratio of mode 3 at different injection strengths as a function of $\Phi_3$ when we replace the optical feedback phase $\omega_3 \tau$ of mode 3 by $\omega_3 \tau + \Phi_3$ in Eq. (1). (b) Dominant mode ratio of mode 3 at different injection strengths as a function of $\Delta f_3$.

*3.2 Control of dominant mode ratio*

In this subsection, we evaluate the effect of chaotic mode-competition dynamics by changing the optical feedback and injection phases. We replaced the term for the optical feedback phase of mode 3 in Eq. (1) from $\omega_3 \tau$ to $\omega_3 \tau + \Phi_3$, where $\Phi_3$ is the phase shift and an optical injection signal to mode 3 is applied. Figure 4(a) shows the dominant mode ratios of mode 3 as a function of $\Phi_3$ at different optical injection strengths for mode 3 ($\kappa_{inj,3}$). The initial optical frequency detuning between the injection light and mode 3 was fixed at $\Delta f_3 = -4.0$ GHz. The dominant mode ratio of mode 3 changed as $\Phi_3$ changed for different $\kappa_{inj,3}$. Particularly, the maximum dominant-mode ratio was observed near $\Phi_3 = 0$ and $2\pi$, whereas the minimum value was obtained near $\Phi_3 = \pi$. Thus, the dominant mode ratio strongly depends on the optical feedback phase $\Phi_3$. From these results, we consider that the difference in the characteristics of the dominant mode ratio in Fig. 3 is due to the difference in the optical feedback phases between the modes. The optical feedback phase for each longitudinal mode is not matched in a multimode semiconductor laser because of the different optical frequencies (wavelengths).

We propose a method to compensate for the differences in the optical feedback phases among the modes by adjusting the initial optical frequency detuning because it is difficult to precisely adjust the optical feedback phase for each mode in the experiment. Both the optical feedback and injection phases affect the dynamics of a semiconductor laser with optical feedback and injection [27,28]. We fixed the optical feedback phase of mode 3 (i.e., $\Phi_3 = 0$) and changed the initial optical frequency detuning of mode 3 ($\Delta f_3$). Figure 4(b) shows the dominant mode ratio of mode 3 as a function of $\Delta f_3$ at different $\kappa_{inj,3}$. Comparing Fig. 4(b) with Fig. 4(a), very similar characteristics of the dominant mode ratio were obtained within the range of 0.1 GHz of $\Delta f_3$. Here, 0.1 GHz corresponds to the inverse of the round-trip time $\tau =10.01$ ns in the external cavity for optical feedback (i.e., the inverse of the feedback delay time). Therefore, similar characteristics of the dominant mode ratio can be obtained by changing $\Delta f_3$ instead of $\Phi_3$.

From these results, the difference in the characteristics of the dominant mode ratio can be compensated by adjusting $\Delta f_3$. The difference in the optical feedback phase between neighboring modes can be described as $2\pi\Delta\nu\tau$, where $\Delta\nu$ is the longitudinal mode spacing. We can compensate for the optical frequency detuning of mode $m$ to match the characteristics of the central mode $m_c$ as follows:

$$\Phi_{adjust,m} = 2\pi(m_c - m)\Delta\nu\tau \quad (-\pi \leq \Phi_{adjust,m} \leq \pi) \quad (5)$$

$$\Delta f_m = \Delta f_{m_c} + \frac{1}{\tau}\frac{1}{2\pi}\Phi_{adjust,m} \quad (6)$$

where Eq. (5) represents the phase shift to match the optical feedback phase of mode $m$ with that of the central mode $m_c$. In Eq. (6), the phase shift obtained from Eq. (5) was converted to a frequency shift in the range of $1/\tau$ and added to the initial optical frequency detuning.

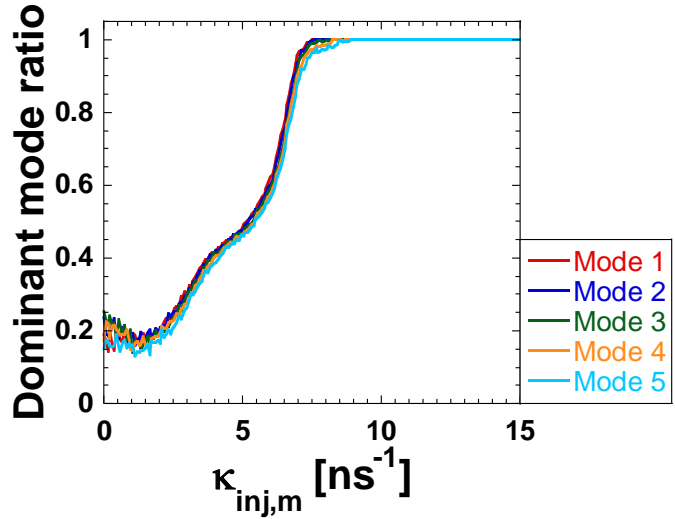

Fig. 5. Comparison of dominant mode ratios of mode $m$ as a function of optical injection strength for mode $m$ when optical injection is applied for only mode $m$ under optical feedback. Initial optical frequency detuning for each mode is adjusted using Eqs. (5) and (6).

The initial optical frequency detuning of the central mode (mode 3) is set to $\Delta f_3 = -4.0$ GHz, and the initial optical frequency detuning $\Delta f_i$ of mode $m$ is adjusted using Eqs. (5) and (6), respectively. Particularly, $\Delta f_i$ of mode 1, 2, 3, 4, and 5 is set to −3.951, −3.975, −4.000, −4.025, and −4.049 GHz, respectively. Figure 5 shows the dominant mode ratio for mode $m$ as the optical injection strength of mode $m$ ($\kappa_{inj,m}$) was changed under the adjustment of $\Delta f_m$. The characteristics of the change in the dominant mode ratio were almost the same for all the modes.

Therefore, the differences in the characteristics of the dominant mode ratio in terms of $\kappa_{inj,m}$ can be compensated by adjusting $\Delta f_m$ using Eqs. (5) and (6), respectively.

The difference in the characteristics of the dominant mode ratio among the modes can be explained by the distribution of steady-state solutions in the phase space (see Appendix for details).

## 4. Parameter dependence

### 4.1 Parameter dependence of dominant mode ratio

In the previous section, we showed that the change in the dominant mode strongly depends on the optical injection strength and initial optical frequency detuning. In this section, we evaluate the characteristics of the dominant-mode ratio when the optical injection strength and initial optical frequency detuning are systematically changed. Figure 6(a) shows two-dimensional (2D) maps of the dominant mode ratio when the optical injection strength for mode 3 ($\kappa_{inj,3}$) and initial optical frequency detuning for mode 3 ($\Delta f_3$) are changed simultaneously. The value of $\kappa_{inj,3}$ required for a large dominant-mode ratio increases when $\Delta f_3$ is far from 0 GHz, which indicates that external light with an optical frequency closer to the longitudinal mode can easily excite the dominant mode. For negative $\Delta f_3$, the dominant mode ratio is reduced in the region where $\kappa_{inj,3}$ is small (blue region), whereas this region is smaller for positive $\Delta f_3$. Thus, the asymmetric characteristics with respect to $\Delta f_3 = 0$ GHz result from the characteristics of semiconductor lasers with a nonzero $\alpha$ parameter [12].

Figure 6(b) shows an enlarged view of Fig. 6(a) in the range of $\Delta f_3$ from −4.0 to −3.0 GHz to observe the fine structure of Fig. 6(a). In Fig. 6(b), the dominant mode is changed periodically by changing $\Delta f_3$ with a frequency interval of 0.1 GHz, which corresponds to the inverse of the feedback delay time $1/\tau$. The dominant mode ratio is significantly changed for small changes in $\Delta f_3$ even if $\kappa_{inj,3}$ has the same value. This periodic structure results from the optical phase shift between optical feedback and injection owing to the change in $\Delta f_3$, as described in the previous section.

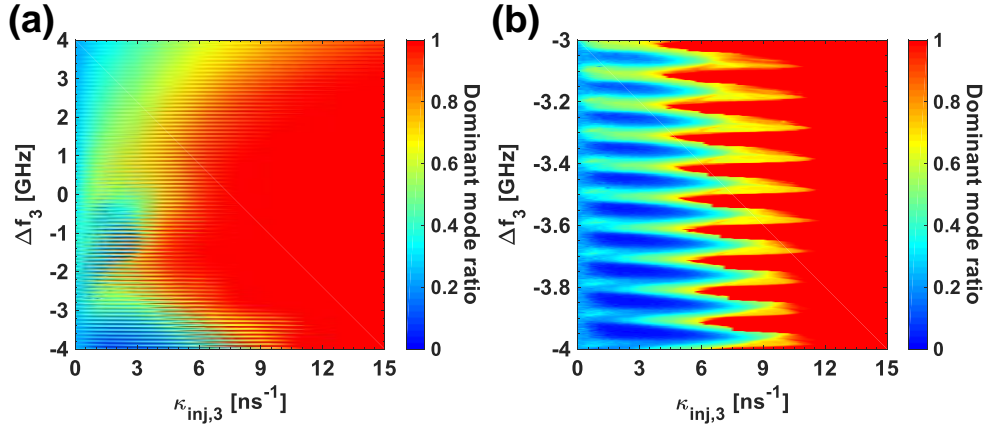

Fig. 6. Two-dimensional maps of (a) dominant mode ratio for mode 3 and (b) its enlarged view. Horizontal axis represents optical injection strength $\kappa_{inj,3}$ for mode 3 and vertical axis represents initial optical frequency detuning $\Delta f_3$ for mode 3.

### 4.2 Relationship between dominant mode ratio and injection locking

In this subsection, we evaluate the relationship between the characteristics of the dominant-mode ratio and injection locking. Figure 7(a) shows the 2D maps of the dominant mode ratio of mode 3 as $\kappa_{inj,3}$ and $\Delta f_3$ were changed in wider ranges than in Fig. 6(a). The region of the large dominant mode ratio of 1 (red region) increased as $\kappa_{inj,3}$ increased. The characteristics of the dominant-mode ratio are asymmetric for $\Delta f_3$. In wide regions of positive $\Delta f_3$, the dominant mode ratio ranges between 0.4 and 0.8. In contrast, the dominant mode ratio is close to zero, and mode 3 is perfectly suppressed in wide regions of negative $\Delta f_3$.

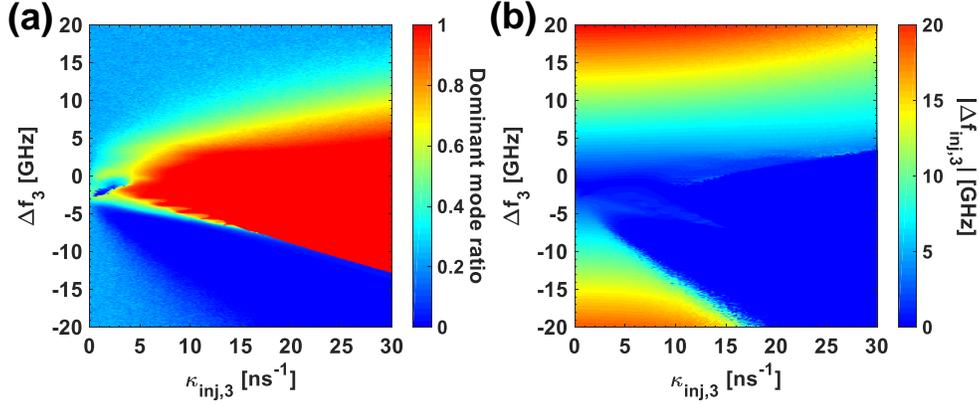

Fig. 7. (a) Two-dimensional map of dominant mode ratio for mode 3 (expanded view of Fig. 6(a)). (b) Absolute value of actual optical frequency detuning for mode 3. Horizontal axis represents optical injection strength $\kappa_{inj,3}$ for mode 3 and vertical axis represents initial optical frequency detuning $\Delta f_3$ for mode 3.

We calculated the actual optical frequency detuning $\Delta f_{inj,3}$ for mode 3 under optical injection. $\Delta f_{inj,3}$ is described using the initial optical frequency detuning $\Delta f_3$ and the difference in the phase changes between the injected light and mode 3 is as follows [12]:

$$\Delta f_{inj,3} = \Delta f_3 + \frac{1}{2\pi}\left[\frac{d\phi_{3,inj}(t)}{dt} - \frac{d\phi_3(t)}{dt}\right]_T \quad (7)$$

where $\phi_{3,inj}(t)$ represents the phase of the injected light and $\phi_3(t)$ represents the phase of mode 3. $[\ ]_T$ represents the average over $T$ ($T$ was set to 20 μs). The phase is calculated from the equation $\phi(t) = \tan^{-1}(E_{im}(t) / E_{re}(t))$, where $E_{re}(t)$ and $E_{im}(t)$ are the real and imaginary parts of the complex electric field amplitude of the laser output, respectively. We set the phase change of $\phi_{3,inj}(t)$ to zero because the injection light was in a steady state. Injection locking is achieved under the condition $\Delta f_{inj,3} \approx 0$ [12].

Figure 7(b) shows the absolute value of the actual optical frequency detuning of mode 3 $|\Delta f_{inj,3}|$ under optical injection. In wide regions of positive $\Delta f_3$, $|\Delta f_{inj,3}|$ increases as $\Delta f_3$ increases from 0 GHz and injection locking does not occur. In contrast, $|\Delta f_{inj,3}|$ approaches 0 GHz and injection locking occurs in wide regions of negative $\Delta f_3$ as $\kappa_{inj,3}$ increases. However, the injection locking range (blue region) in Fig. 7(b) does not directly correspond to the region with a large dominant mode ratio (red region) in Fig. 7(a).

Figure 8 summarizes the relationship between the injection locking range and region of the large dominant-mode ratio by comparing Figs. 7(a) and 7(b). We define $|\Delta f_{inj,3}| \leq 0.001$ GHz as the injection-locking range, as shown in Fig. 7(b). Figure 8 is categorized into four regions: (a) the dominant mode ratio is 1 and injection locking is achieved (blue region in Fig. 8); (b) the dominant mode ratio is 1 and injection locking is not achieved (red region);, (c) the dominant mode ratio is not 1, injection locking is achieved (light green region); and (d) the dominant mode ratio is not 1, and injection locking is not achieved (purple region).

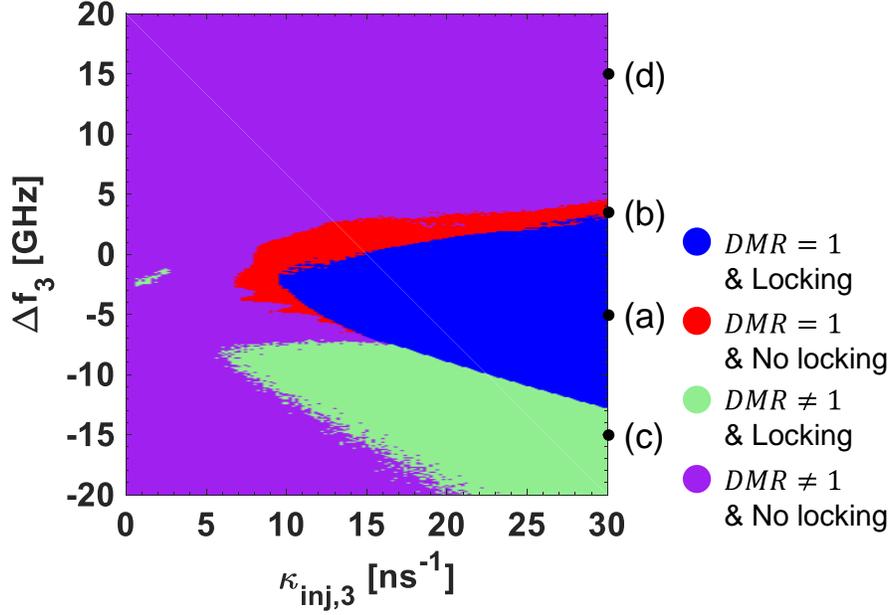

Fig. 8. Two-dimensional map of the dominant mode ratio of 1 and injection locking range. (Blue) dominant mode ratio is 1 and injection locking is achieved; (red) dominant mode ratio is 1 and injection locking is not achieved; (light green) dominant mode ratio is not 1 and injection locking is achieved; and (purple) dominant mode ratio is not 1 and injection locking is not achieved. (a)-(d) correspond to the temporal dynamics shown in Fig. 9.

To show the dynamics in the four regions of Fig. 8, we evaluated the temporal waveforms of the modal intensities and short-term optical-frequency detuning under optical injection for mode 3. The dynamics of short-term optical frequency detuning for mode 3 ($\Delta f_{inj,3}(t)$) is described as follows:

$$\Delta f_{inj,3}(t) = \Delta f_3 + \frac{1}{2\pi}\left[\frac{\phi_{3,inj}\left(t+\frac{\tau_\phi}{2}\right) - \phi_{3,inj}\left(t-\frac{\tau_\phi}{2}\right)}{\tau_\phi} - \frac{\phi_3\left(t+\frac{\tau_\phi}{2}\right) - \phi_3\left(t-\frac{\tau_\phi}{2}\right)}{\tau_\phi}\right] \quad (8)$$

where $\tau_\phi$ is the duration for averaging the optical phase shift and $\tau_\phi$ is set to 0.1 ns to observe fast frequency dynamics. We set $\phi_{3,inj}(t + \tau_\phi/2) - \phi_{3,inj}(t - \tau_\phi/2)$ to zero because the injection light is in a steady state.

Figure 9 shows the temporal waveforms of the five modal intensities and short-term optical-frequency detuning for mode 3 $\Delta f_{inj,3}(t)$ in the presence of optical injection ($\kappa_{inj,3}$ = 30.0 ns$^{-1}$) for different $\Delta f_3$. The dynamics of Figs. 9(a)–(d) correspond to examples of the four regions indicated by (a)–(d) in Fig. 8.

Figure 9(a) shows the temporal waveforms of the five modal intensities and $\Delta f_{inj,3}(t)$ for $\Delta f_3$ = −5.0 GHz, which corresponds to the blue region in Fig. 8. The temporal oscillations of all the modes were stabilized in steady states. Mode 3 has the maximum intensity, whereas the other modes have zero intensity. Therefore, mode 3 becomes the dominant mode. Moreover, $\Delta f_{inj,3}(t)$ is stabilized at 0 GHz, and perfect injection locking is achieved.

Figure 9(b) shows the temporal waveforms of the five modal intensities and $\Delta f_{inj,3}(t)$ for $\Delta f_3$ = 3.5 GHz, which corresponds to the red region in Fig. 8. The temporal waveform of mode 3 oscillates quasi-periodically, and the other modes are perfectly stabilized with zero intensities

(no oscillations). Therefore, mode 3 is always the dominant one even though it exhibits quasi-periodic oscillations. Additionally, $\Delta f_{inj,3}(t)$ oscillates between −2 and 8 GHz, and injection locking does not occur, even on average. Therefore, a large dominant-mode ratio can be achieved even without injection locking in the red region of Fig. 8.

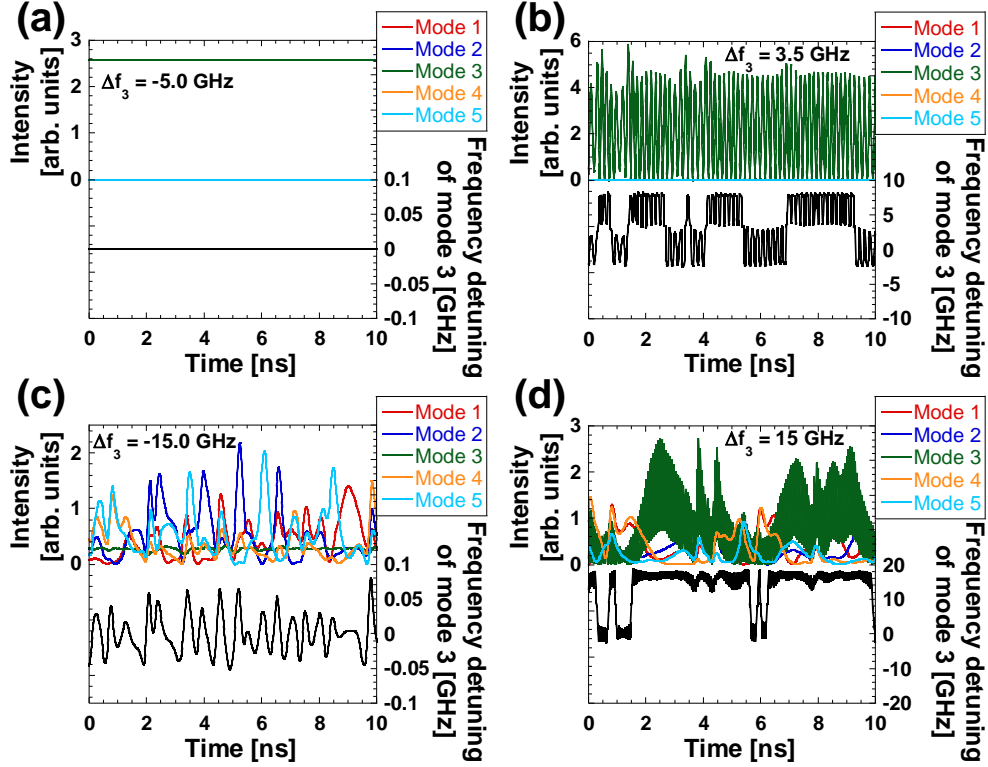

Fig. 9. Temporal waveforms for different initial optical frequency detuning $\Delta f_3$ for mode 3 at $\kappa_{inj,3} = 30.0$ ns$^{-1}$. (upper) Five modal intensities and (lower) actual optical frequency detuning of mode 3. (a) $\Delta f_3 = -5.0$ GHz. (b) $\Delta f_3 = 3.5$ GHz. (c) $\Delta f_3 = -5.0$ GHz. (d) $\Delta f_3 = 15.0$ GHz. (a)-(d) correspond to the regions shown in Fig. 8.

Figure 9(c) shows the temporal waveforms of the five modal intensities and $\Delta f_{inj,3}(t)$ for $\Delta f_3 = -15.0$ GHz, which corresponds to the light green region in Fig. 8. Only the temporal waveform of mode 3 was stabilized with small fluctuations by optical injection, and the other modes exhibited large chaotic oscillations. Therefore, mode 3 was not the dominant one. Moreover, $\Delta f_{inj,3}(t)$ fluctuates chaotically around 0 GHz within the range of ±0.1 GHz; however, the mean of $\Delta f_{inj,3}(t)$ is close to 0 GHz, where injection locking seems to be achieved on average. In fact, $\Delta f_{inj,3}(t)$ fluctuates slightly because of the chaotic mode-competition dynamics from the other modes, and injection locking is achieved on average. However, the dominant mode ratio of mode 3 does not become 1 because the temporal dynamics of mode 3 are almost stabilized, whereas the other modes fluctuate chaotically.

Figure 9(d) shows the temporal waveforms of the five modal intensities and $\Delta f_{inj,3}(t)$ for $\Delta f_3 = 15.0$ GHz, which corresponds to the purple region in Fig. 8. All the modes oscillate chaotically, and the dominant mode ratio is not 1. Mode 3 oscillates with faster frequencies than those of the other modes owing to the beat frequency of $\Delta f_3$ (= 15.0 GHz). Additionally, $\Delta f_{inj,3}(t)$ remains near 17 GHz and occasionally moves to 0 GHz, indicating that injection locking is not achieved on average.

From these results, we deduce that the region of the dominant mode ratio of 1 did not perfectly match the injection locking range. A dominant mode ratio of 1 under injection locking can be obtained only in the blue region of Fig. 8. However, a dominant mode ratio of 1 can still be achieved without injection locking in the red region of Fig. 8, where the intensity and optical frequency of mode 3 oscillate periodically (or quasi-periodically or chaotically) and the other modes are perfectly stabilized. In contrast, the dominant mode ratio cannot reach 1 even though injection locking is achieved on average in the green-light region of Fig. 8, where only mode 3 is suppressed and the other modes fluctuate chaotically. The optical frequency of mode 3 fluctuates slightly around 0 GHz, which indicates that injection locking is achieved only on average. Finally, a dominant mode ratio of 1 was not obtained, and injection locking was not achieved in the purple region of Fig. 8. Therefore, the relationship between the region of the dominant mode ratio of 1 and the injection locking range is not straightforward in a multimode semiconductor laser with optical feedback and injection, unlike the case of a single-mode semiconductor laser [12].

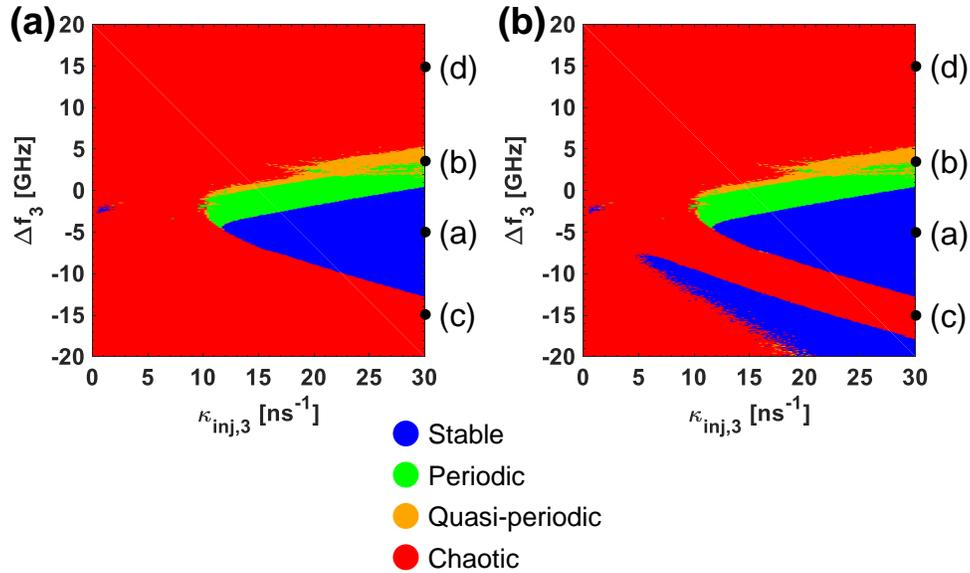

Fig. 10. Two-dimensional maps of the temporal dynamics of (a) total intensity and (b) modal intensity of mode 3 as $\kappa_{inj,3}$ and $\Delta f_3$ are changed. (a)-(d) correspond to the temporal dynamics shown in Fig. 9.

Finally, we evaluated the temporal dynamics of different values of $\kappa_{inj,3}$ and $\Delta f_3$. Figure 10(a) shows the 2D map of the temporal dynamics of the total intensity in the multimode semiconductor laser as $\kappa_{inj,3}$ and $\Delta f_3$ are changed simultaneously. A steady state is observed in the blue triangle region for a wide region of negative $\Delta f_3$ and large $\kappa_{inj,3}$. The periodic (green region) and quasi-periodic (orange region) oscillations are located around the upper side of the blue triangular region of the steady state (near zero $\Delta f_3$ and large $\kappa_{inj,3}$). The other region corresponds to chaotic oscillations, indicated by the red region in Fig. 10(a), owing to optical feedback. The regions of steady state and periodic oscillations in Fig. 10(a) correspond to the blue region in Fig. 8, where the dominant mode ratio is 1 and injection locking is achieved.

Figure 10(b) shows a 2D map of the temporal dynamics of mode 3 as $\kappa_{inj,3}$ and $\Delta f_3$ are changed. A new region of the steady state (blue region) appears at a large negative $\Delta f_3$ and large $\kappa_{inj,3}$, which is different from the temporal dynamics of the total intensity in Fig. 10(a). This region is included in the light-green region of Fig. 8, where the dominant mode ratio is not 1, even though injection locking is achieved. The dynamics of mode 3 are stabilized under injection locking; however, the dominant mode ratio is not 1, owing to the appearance of the chaotic oscillations of the other modes. Therefore, the difference in the dynamics between the total and modal intensities results in a mismatch between the region of the large dominant mode ratio and the injection locking range, as shown in Fig. 8.

## 5. Conclusions

In this study, we numerically evaluated the chaotic mode-competition dynamics in a multimode semiconductor laser with optical feedback and injection. Chaotic mode-competition dynamics were observed in the longitudinal modes, and one of the longitudinal modes was enhanced by injecting an optical signal with a wavelength similar to that of a single-mode semiconductor laser. The dominant mode was defined as the mode with the maximum intensity, and the dominant-mode ratio for the injected mode increased as the optical injection strength increased. However, the characteristics of the dominant mode ratio in terms of the optical injection strength were different among the modes owing to the difference in the optical feedback phase. We proposed a control method to match these characteristics by adjusting the initial optical frequency detuning between the injection signal and injected mode in the multimode laser. Additionally, we evaluated the relationship between the region of a large dominant mode ratio and the injection locking range. The large dominant-mode ratio region did not match the injection-locking range. This discrepancy results from the complex mode-competition dynamics in multimode semiconductor lasers with both optical feedback and injection.

The control technique of the chaotic mode-competition dynamics in multimode semiconductor lasers could be useful for photonic computing applications in reinforcement learning and reservoir computing as novel photonic AI.

## 6. Appendix

### 6.1 Steady state analysis

In the Appendix section, we explain the influence of the optical feedback phase on the dominant mode ratio using steady-state analysis. The initial optical frequency detuning of 0.1 GHz (= $1/\tau$) corresponds to an optical feedback phase of $2\pi$, as shown in Figs. 4(a) and 4(b). Generally, multiple steady-state solutions appear in the frequency interval corresponding to $1/\tau$ in a single-mode semiconductor laser with optical feedback [12,42]. Thus, the distribution of steady-state solutions may be related to the change in the dominant mode ratio.

Here, we consider steady-state solutions for only one longitudinal mode in a multimode semiconductor laser with optical feedback for simplicity. In this case, the steady-state solutions are almost identical to those of a single-mode semiconductor laser with optical feedback, except for the steady-state solution of the carrier density, which depends on the gain coefficient of each mode. The steady-state solutions with optical feedback for the carrier density $N_s$ and angular frequency $\omega_{s,q}$ of the longitudinal mode $q$ are expressed as follows:

$$\Delta\omega_{s,q} = -\kappa\sqrt{1+\alpha^2}\sin(\Delta\omega_{s,q}\tau + \omega_q\tau + \tan^{-1}\alpha) \tag{9}$$

$$N_s = \frac{\tau_s G_q N_0 + \tau_s/\tau_p + \varepsilon\, N_{th}\, J/J_{th} - 2\kappa\tau_s \cos(\omega_{s,q}\tau)}{\tau_s G_q + \varepsilon} \tag{10}$$

where $\Delta\omega_{s,q}$ indicates $\Delta\omega_{s,q} = \omega_{s,q} - \omega_q$, $\omega_{s,q}$ is the steady-state solution for the angular frequency of the longitudinal mode $q$ with optical feedback, and $\omega_q$ is the angular frequency of mode $q$ without optical feedback. $G_q$ indicates the gain coefficient of mode $q$. The steady-state solutions for the angular frequency ($\omega_{s,q}$) are converted to those for the frequency, that is, $\nu_{s,q} = \omega_{s,q}/2\pi$.

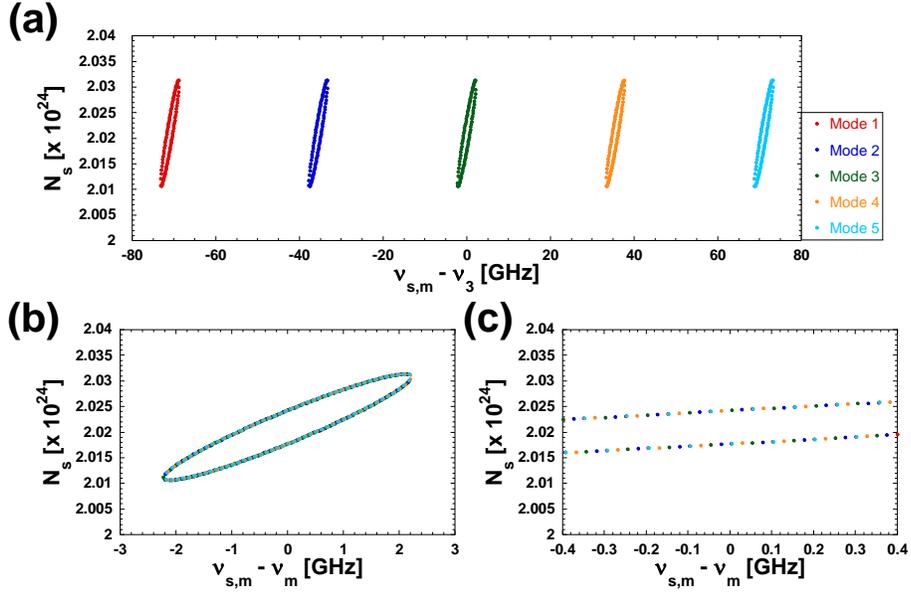

Fig. 11. Steady-state solutions obtained using Eqs. (9) and (10). (a) Distribution of steady-state solutions for five longitudinal modes with respect to mode 3. (b) Distribution of steady-state solutions with respect to each longitudinal mode. Steady-state solutions are overwritten for the five modes. (c) Enlarged view of (b). Steady-state solutions for the five modes are shown with different colors.

Figure 11 shows the steady-state solutions for mode $m$ obtained using Eqs. (9) and (10), respectively. Figure 11(a) shows the distribution of the steady-state solutions for each longitudinal mode with respect to the central mode (mode 3). The distributions were obtained by adding the mode spacing $\nu_m - \nu_3$ to the steady-state solution. Multiple steady-state solutions are elliptically distributed around each of the five longitudinal modes. Figure 11(b) shows the distribution of the steady-state solutions for each longitudinal mode with respect to the modal frequency $\nu_m$ of the longitudinal mode $m$ without optical feedback and injection. The steady-state solutions of all modes are elliptically distributed; however, they are shifted for each longitudinal mode. Figure 11(c) shows an enlarged view of Fig. 11(b). The steady-state solutions for each longitudinal mode were distributed at intervals of 0.1 GHz (corresponding to $1/\tau$); however, they were placed at different frequencies.

Here, the steady-state solutions can be obtained by determining the intersection of $\Delta\omega_{s,q}$ (that is, the left-hand-side term in Eq. (9)) and sinusoidal waves (that is, the right-hand side term in Eq. (9)) [12,42]. The initial phase of the sinusoidal wave on the right-hand side term in Eq. (9) is determined by the term $\omega_q \tau$ and the position of the intersection, and the difference in the optical feedback phase for each mode ($\omega_q \tau$) affects the positions of the steady-state solutions. Particularly, the steady-state solutions of modes 1 and 5 almost overlap in Fig. 11(c) because the feedback phases $\omega_q \tau$ for modes 1 and 5 are very similar. The characteristics of the dominant mode ratios of modes 1 and 5 in terms of the optical injection strength are very similar, as shown in Fig. 3.

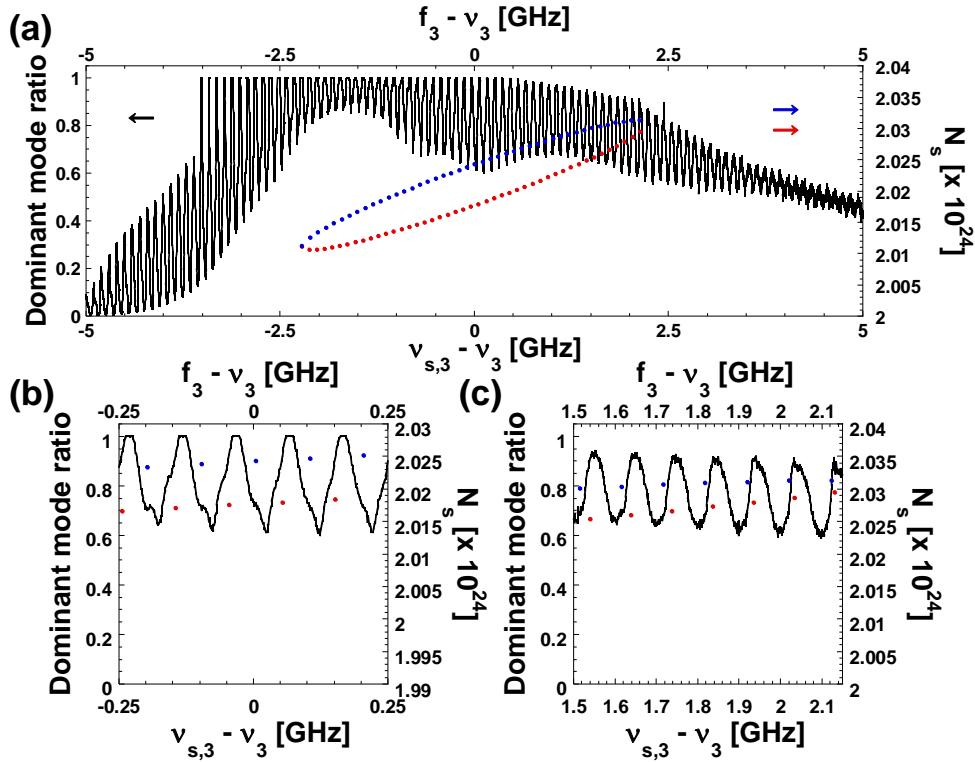

Fig. 12. (a) Dominant mode ratio of mode 3 (black curve) when initial optical frequency detuning for mode 3 is changed and optical injection strength for mode 3 is fixed at $\kappa_{inj,3} = 5.0$ ns$^{-1}$. Distributions of steady-state solutions for mode 3 are also shown (red and blue dots for modes and anti-modes, respectively). (b), (c) Enlarged views of (a) at different frequency ranges.

The steady-state solutions are obtained without optical injection. The steady-state solutions may be changed by the optical injection; however, those with only optical feedback are used as the original dynamics because we interpret the optical injection as a small perturbation to the chaotic multimode semiconductor laser with optical feedback.

Figure 12(a) shows the dominant mode ratio of mode 3 (black curve) when the initial optical frequency detuning for mode 3 ($\Delta f_3$) was changed, and the optical injection strength for mode 3 was fixed at $\kappa_{inj,3} = 5.0$ ns$^{-1}$. The dominant mode ratio changes at different initial optical frequency detunings, although $\kappa_{inj,3}$ is constant. The dominant mode ratio repeatedly changes with a frequency interval of 0.1 GHz for a wide range over ±5.0 GHz, although the steady-state solutions exist only in the range within ±2.3 GHz. In other words, the effect of optical feedback appears even outside the range of steady-state solutions. Figure 12(a) also shows the distributions of the steady-state solutions of Mode 3 obtained from Eqs. (9) and (10) to understand the relationships between the dominant mode ratio and distributions of the steady-state solutions. The lower half of the ellipse (red dots) is known as the external-cavity mode (or mode), which is a stable solution, whereas the upper half of the ellipse (blue dots) is known as the anti-mode, which is an unstable solution [12,42].

Figures 12(b) and 12(c) show enlarged views of Fig. 12(a) for different ranges to understand the relationship between the dominant-mode ratio and distributions of the steady-state solutions. The change in the dominant mode ratio occurs in the frequency interval of 0.1 GHz (= $1/\tau$). It is worth noting that the dominant mode ratio increases at the initial optical frequency detuning near the modes (red dots), whereas the dominant mode ratio decreases at the initial optical frequency detuning near the anti-modes (blue dots). Thus, the dominant mode ratio is affected by the distributions of the steady-state solutions (modes and anti-modes).

We discuss the relationship between the initial optical frequency detuning and steady-state solutions. The relative positions of the modes and anti-modes on the horizontal axis in Fig. 12 (optical frequency detuning) are related to the dominant mode ratio. In Fig. 4, the increase in the optical feedback phase (corresponding to the shift of the steady-state solutions in the negative direction) is equivalent to the shift in the initial optical frequency detuning in the positive direction. Therefore, a change in the initial optical frequency detuning is effective in changing the position of the steady-state solution, which results in a change in the dominant mode ratio.

It has also been reported that the distribution of the modes and anti-modes can be used to control chaos in a semiconductor laser with optical feedback [43]. The dominant mode ratio can be controlled by changing the distribution of the modes and anti-modes in the proposed scheme. This is performed using the method proposed in Section 3.2.

**Funding.** Grants-in-Aid for Scientific Research from the Japan Society for the Promotion of Science (JSPS KAKENHI, Grant No. JP19H00868, JP20K15185, and JP22H05195), JST CREST, Japan (JPMJCR17N2), and the Telecommunications Advancement Foundation.

**Disclosures.** The authors declare no conflicts of interests.

**Data availability.** The data underlying the results presented in this paper are not publicly available at this time but may be obtained from the authors upon reasonable request.